\documentclass[preprint]{aastex}

\usepackage{graphicx}
\usepackage{amssymb}
\usepackage{amsmath}
\usepackage{color}
\usepackage{bm}
\usepackage{xspace}
\usepackage{pgffor}
\usepackage{upgreek}
\usepackage{nicefrac}
\usepackage[normalem]{ulem}
\usepackage[single=true]{acro}
\usepackage{wrapfig}

\newcommand{\be}{\begin{equation}}
\newcommand{\ee}{\end{equation}}
\newcommand{\nn}{\mbox{} \nonumber \\ \mbox{} }
\newcommand{\ba}{\begin{eqnarray}}
\newcommand{\ea}{\end{eqnarray}}
\newcommand{\om}{\omega}
\newcommand{\Alfven}{ Alfv\'{e}n }

\newcommand{\Bf}{{magnetic field}}
\newcommand{\Ef}{{electric field}}
\newcommand{\Bfs}{{magnetic fields}}
\newcommand{\NS}{neutron star}
\newcommand{\ms}{magnetosphere}
\newcommand{\mss}{magnetospheres}
\newcommand{\NSs}{{neutron stars}}

\newcommand{\EM}{electromagnetic}

\newcommand\eg{\textit{e.g.}}

\newcommand\lo{\mathrel{\raise.3ex\hbox{$<$}\mkern-14mu\lower0.6ex\hbox{$\sim$}}}
\newcommand\go{\mathrel{\raise.3ex\hbox{$>$}\mkern-14mu\lower0.6ex\hbox{$\sim$}}}

\begin{document}

 \title{FRB-periodicity: mild pulsars in tight   O/B-star binaries}
 
\author{Maxim Lyutikov$^{1}$,
 Maxim V. Barkov$^{1,2}$,
  Dimitrios Giannios$^1$}

\affil{
$^1$ Department of Physics, Purdue University, 525 Northwestern Avenue, West Lafayette, IN 47907-2036, USA \\
$^2$ Astrophysical Big Bang Laboratory, RIKEN, 2-1 Hirosawa, Wako, Saitama 351-0198, Japan}


\date{Received/Accepted}

\begin{abstract}
{Periodicities observed in two Fast Radio Burst (FRB) sources (16 days  in  FRB 180916.J0158+65 and  160 days in FRB 121102) are consistent with that of  
tight,  stellar mass  binary systems.}  In the case of   FRB 180916.J0158+65 the primary is  an early OB-type star  with  
mass loss rate $\dot{M} \sim 10^{-8}- 10^{-7} M_\odot$ yr$^{-1}$, and the secondary a neutron star. The observed  periodicity is not intrinsic to the  
FRB's source, but is due to the orbital phase-dependent  modulation of  the  absorption conditions in the massive star's wind. 
The observed relatively narrow FRB activity window implies  that the primary's wind dynamically dominates that of the pulsar, 
$\eta = L_{sd}/(\dot{M} v_w c) \leq 1$, where $L_{sd} $ is pulsar spin-down, $\dot{M}$ is the primary's wind mass loss rate and $v_w$ is its velocity. 
The condition $\eta \leq 1$  requires  mildly powerful pulsar   with $L_{sd} \lesssim  10^{37}$ erg $s^{-1}$.    The observations are consistent 
with magnetically-powered radio emission originating  in the \mss\ of  young, strongly magnetized {\NS}s, the classical  magnetars.
\end{abstract}


\maketitle

\section{FRB periodicity  due to the orbital motion in  O/B-NS binary}

\subsection{Observations and outline of  the model}

{The CHIME collaboration announced a $P=16$~day periodicity from   FRB 180916.J0158+65 \citep{2020arXiv200110275T}.  Analysis of date from 76-m Lovell telescope on the  original repeater FRB~121102 also indicated periodicity of $P \sim 160$ days \citep{2020arXiv200303596R}.
These are  important  observations which sheds light on the origin of FRBs, as we discuss in the  present Letter. Below we concentrate on the case of   FRB 180916.J0158+65.
}

{
The observed periodicity is most likely due to the orbital motion of a binary system. (In Appendix \ref{Unlikely} we discuss an unlikely periodicity due to geodetic precession in extremely tight binary. Free precession models were also proposed \cite{2020arXiv200204595L,2020ApJ...892L..15Z}). 
}

 The orbital semi-major axis  for the case of  FRB 180916.J0158+65 evaluates to 
\ba &&
P = 2\pi \sqrt{ \frac{a^3}{G M_\odot (m_{PSR}+ m_{MS})}}
\nn &&
a = 4 \times 10^{12} m_{\rm tot,1}^{1/3} P_{1.2}^{2/3} \; {\rm cm},
\label{ac} 
\ea
where $m_{PSR}$ is   (the presumed)  \NS's mass and $m_{MS}$ is  the primary's  mass  in Solar masses and $m_{\rm tot}=(m_{PSR}+ m_{MS})$. (We refer to the the O/B star  as ``a primary", while to the \NS\  as a  ''companion".)  In this paper we use the  notation $A_x=A/10^x$ and $m_{\rm tot}$ is measured in Solar masses $M_\odot$, P is in days.

\begin{figure}[h!]
 \begin{center}
\includegraphics[width=.44\linewidth]{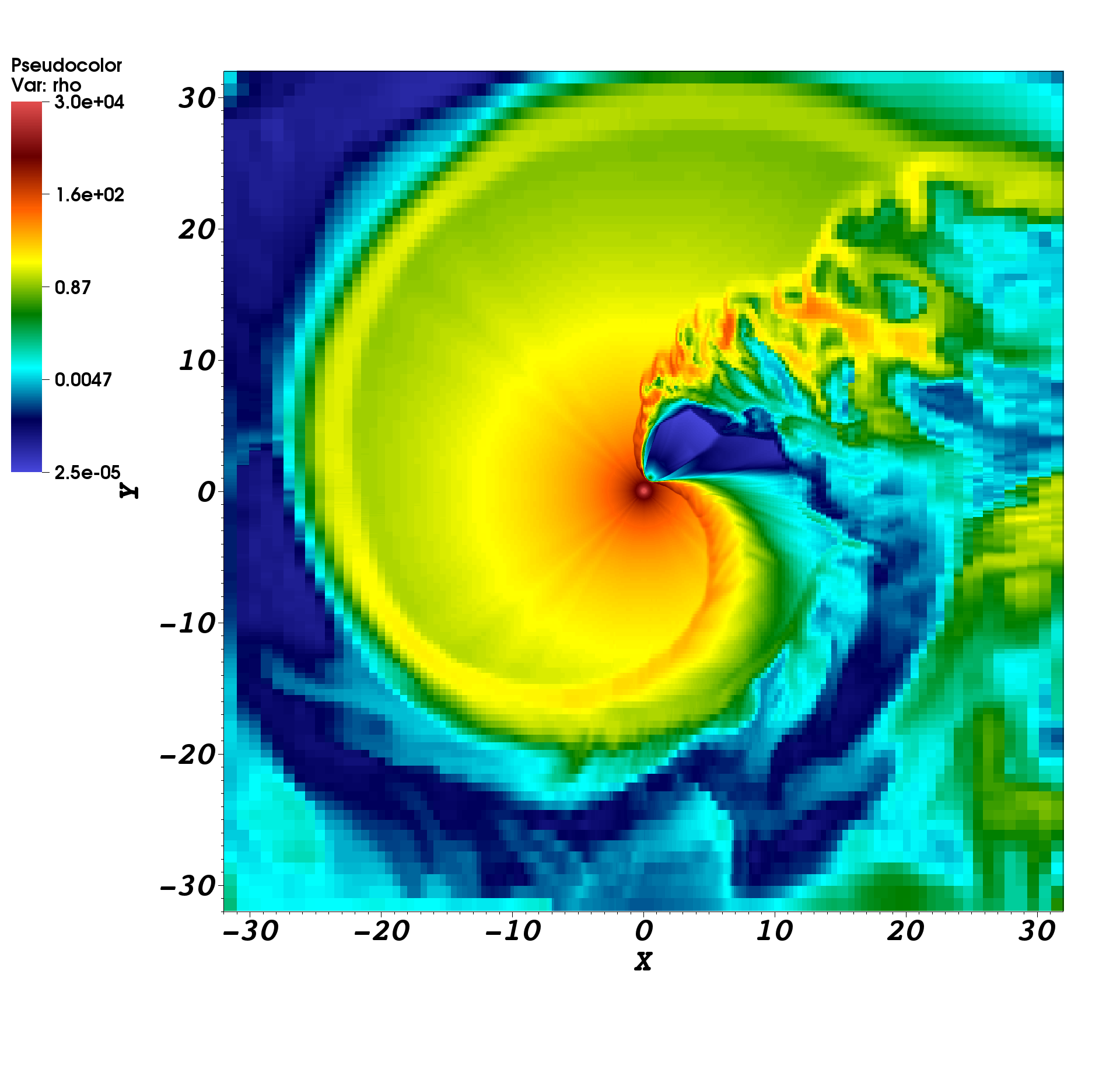}
\includegraphics[width=.50\linewidth]{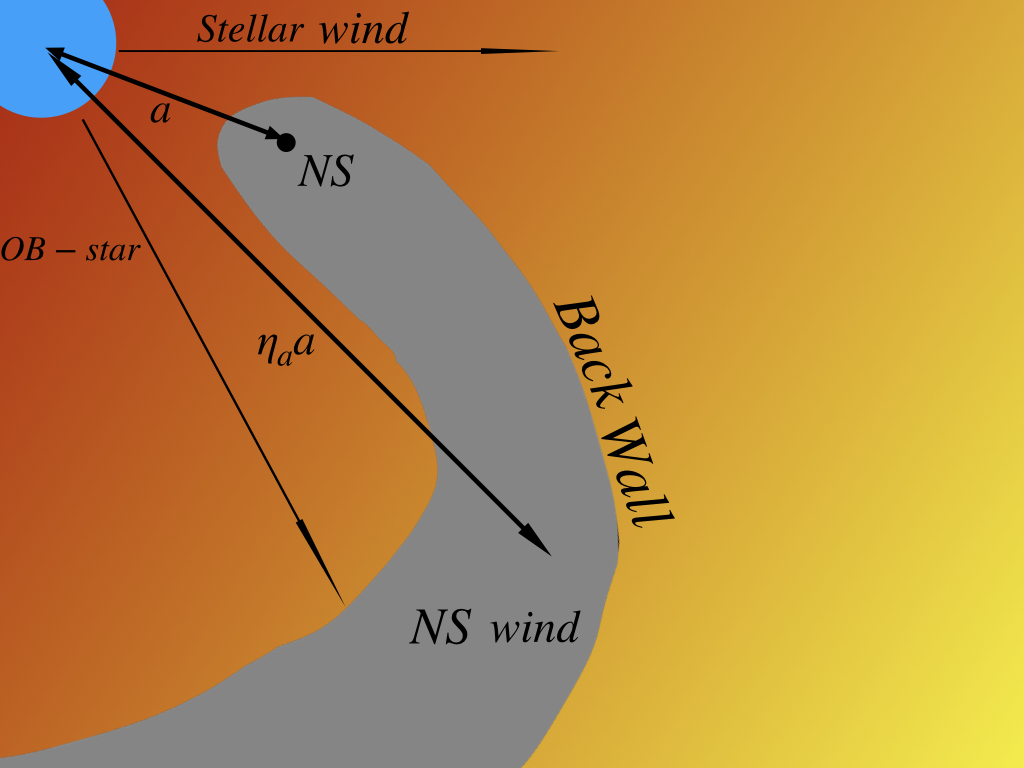}
\end{center}
\caption{ Left panel:  reprocessed simulations of \protect\cite{2015A&A...577A..89B} of interacting winds of B-star and pulsar's. Right panel: artistic image of the interacting winds.  The pulsar wind creates a narrow low density cavity that extends to distances much larger than the orbital separation by a factor $\eta_a \sim 10-30$.  At that point the density of the primary's wind is lower by $\eta_a^2$,   the absorption coefficient is reduced by  $\eta_a^4$, and optical depth by  $\eta_a^3$.}
\label{WIndow}
\end{figure}

Both the pulsar,  {\it loci} of the FRB, and the primary produce winds: a relativistic wind by the pulsar and wind with velocity $v_w \sim$ few $10^3$ km s$^{-1}$ by main-sequence star \citep{2001A&A...369..574V}. We hypothesize  that the observed periodicity is due to absorption of the FRB pulses in the primary's wind,  see Fig. \ref{WIndow}.

The interacting pulsar's and the primary's  winds  create a conically shaped cavity around the less powerful source. The primary's   
winds can be highly optically thick at radio waves  due to free-free absorption, while the relativistic pulsar wind is, basically, transparent to radio waves.
A transparent cone-like zone is, therefore, created behind the pulsar, modified into spiral structure by the orbital motion. 

The radio waves propagating within this spiral structure do not experience free-free absorption. After they enter the primary's wind, at larger distance, the  wind's density, and the corresponding absorption coefficient, are substantially reduced. Thus,  the dynamics of interacting winds creates transparency windows when the observer sees the pulsar through the spiral structure  \citep{2011A&A...535A..20B,2012A&A...544A..59B,2015A&A...577A..89B}. 
 
Since the active window is observed to be less than 50\% of the orbital period, {\it  the primary' s wind 
should dominate over that of the pulsar}. This  requires  that the momentum parameter $\eta $  is less than unity,
\be
\eta =\frac{L_{sd}}{\dot{M} v_w c }\leq 1.
\label{eta}
\ee
\citep[This regime is opposite to the case of Black Widow-type pulsar binaries][]{1988Natur.333..237F}.

The typical opening of the transparency wedge behind a pulsar as seen in simulations of \cite{2015A&A...577A..89B} is $\sim 15-20 ^\circ$; 
this is weakly dependent on the momentum parameter as long as $\eta \lesssim 0.3$. This is consistent with the active phase observed in FRB 180916.J0158+65. 

Simulations of  \cite{2015A&A...577A..89B}, see also Fig. \ref{WIndow},
demonstrate 
 that the wind cavity can extend out to 10-30 times the binary separation radius. 
As a result, the cavity created by the pulsar's wind can reduce the absorption optical depth by a factor as large as $10^3$:
 if at the location of the pulsar the primary's wind is still moderately, $\tau \leq 10^3$, optically thick to infinity, and the  observer's line of sight lies close to the equatorial plane, within  $\sim 15-20 ^\circ$, then the source turns transparent to radio emission for part of the orbit.

\subsection{Free-free absorption in the primary's wind}
\label{{model}}
 Below we make the numerical estimates of the proposed model.
We parametrize the location  of the ``back wall" of the cavity as $\eta_a a$ ($a$ is orbital separation), and normalize the numerical estimates to  $\eta_a = 10^{1.5} \eta_{a,1.5}$.

The free-free optical depth  form the location of the ``back wall" of the cavity  to infinity should be of the order of unity for radio waves to escape.
Using the  free-free-absorption coefficient  $\kappa_{ff}$ \citep{1999acfp.book.....L}, assuming temperature $T=10^4$ K, and  scaling the ``back wall"  radius with the orbital separation (\ref{ac}), we find
\ba &&
 \kappa_{ff}  = 3.28 \times 10^{-7} T_4^{-1.35} \nu_{1 GHz}^{-2.1} \frac{EM}{pc}
\nn &&
\dot{M} = 4\pi n m_p ( \eta_a a) ^2 v_w 
\nn &&
\tau _{ff}  \sim   5\times10^{-3} \; \eta_{a,1.5}^{-3}   \dot{M}_{-7.5}^2  T_4^{-1.35} \nu_{1 GHz}^{-2.1} m_{\rm tot,1} v_{w,8.5}^{-2},
\label{Mdot}
\ea
here $EM=\int_a^{\infty} n^2 dr$ is emission measure.
So, a mass loss rate of $ \dot{M} \sim 3\times 10 ^ {-8} M_\odot$/yr can give substantial optical depth to free-free absorption for a source at distance $a$ but transparent conditions when the source is observed through the
 ``back wall" of a cavity that extends  to $\sim 30  a$.

Using mass loss rate $\dot{M}$ from (\ref{Mdot}) and   requiring  the primary's momentum dominance in the wind-wind interaction, we estimate the pulsar spindown luminosity 
\be
L_{sd} \leq  5 \times 10^{36} \eta_{-0.5}  \dot{M}_{-7.5} v_{w,8.5}\; {\rm erg \; s}^{-1}.
\ee
Thus, the pulsar can produce mildly strong winds.


 The model limits the mass loss  rate  of the main sequence star both from below (the wind-wind interaction should be dominated by the Main Sequence star), and from above 
(too powerful winds are optically thick to free-free absorption out to very large distances). 
Early 
B-type stars,  and late O-type stars  fit the required mass loss range.  Late B-type stars have insufficiently strong winds, while earlier O-type stars remain optically thick to large distances \citep{2001A&A...369..574V,2014A&A...564A..70K}.

\subsection{Nonlinear propagation effects}

Strong radio waves can affect non-linearly  the free-free absorption coefficient in the wind \citep{2019arXiv191212241L}. Qualitatively, strong \EM\ wave in weakly magnetized plasma induces electron's oscillations with momentum
\ba &&
p_\perp \sim a_0 m_e c
\nn && 
a_0 = \frac{e E} {m_e c \om},
\label{a0}
\ea
 where $a_0$ is the laser intensity parameter, $E$ is the \Ef\ in the wave \citep{1975OISNP...1.....A}. If the quiver energy  $\sim a_0^2 m_e c^2$ is larger than temperature, $a_0 \geq \sqrt{T/(m_e c^2)}$,  then the motion of the electrons is determined by the wave itself, not the temperature.

For typical parameters of   FRB 180916.J0158+65 \citep{2020arXiv200110275T} (total fluence ${\cal{ F} } = 10 $~Jy~msec,  duration $\tau =  2$ msec and distance
$d_{FRB} =150 $ Mpc)  the luminosity and the  total energy in a single burst evaluate to 

\ba &&
L_{FRB} = 4 \pi  \frac{d_{FRB}^2 { \cal{ F} }}{\tau} = 3 \times 10^{40} \,\; {\rm erg\,  s}^{-1}.
\nn &&
E_{FRB} = 4 \pi  d_{FRB}^2 { \cal{ F} } = 6 \times 10^{37} \,\; {\rm erg.}
\ea
The nonlinearity parameter is then 
\be
a_0 = \frac{ e \sqrt{ { \cal{ F} } } }{ a m_e c^{3/2} \sqrt{\tau} \om} \approx 1. 
\ee
Thus, the FRB can  heat  the  plasma to relativistic energies! As a result, a sufficiently strong pulse can propagate much further than expected from the linear theory, 
heating up the plasma and decreasing the absorption.  But the radio pulse  does not have enough energy to escape, as we discuss next.   

As the particles are heated by the wave, the plasma absorption coefficient $\kappa_{ff}$ decreases, allowing a wave to propagate further. 
The total amount of the  material  the FRB heats is 
\be
M_h \sim \frac{ m_p E_{FRB} }{a_0^2 m_e c^2} = 10^{20}\; {\rm g}.
\ee
This huge amount  of material heated to relativistic electron temperatures is still much smaller than contained in the wind within the orbit
\be
M_w \approx \frac{a}{v_w} \dot{M} = 3 \times 10^{22}   \dot{M}_{-7.5} \; {\rm g}
\ee
or 
\be
\frac{M_h}{M_w} \approx 3\times 10^{-3}   \dot{M}_{-7.5}^{-1}.
\ee

Thus, a leading part of the FRB   heats a part of the wind to relativistic energies, which makes it more transparent to the trailing part of the emission. 
But the FRB energetics is not sufficient to make the wind fully transparent, so the estimates in \S \ref{model} remain valid. Also, the momentum implanted by the FRB pulse, 
$P_{FRB} \sim E_{FRB}/c \sim 2 \times 10^{27}$  g cm s$^{-1}$ is much smaller that the momentum of the wind within orbital radius,  $P_w \approx a \dot{M} 
= 7 \times 10^{30} \dot{M}_{-7.5} $  g cm s$^{-1}$; the FRB pulse does not disturb the overall outflow. Moreover, the kinetic energy of the stellar wind $E_{w,kin}=M_w v_w^2/2 \approx 10^{39}$~ergs 
is much larger compared to the FRB energy: $E_{FRB}/E_{w,kin} \sim 1/20$, so FRB impact on the flow dynamics can be neglected.  

Non-linear effects are strongly suppressed by the \Bf\ \citep{2019arXiv190103260L,2020arXiv200109210L}. If the cyclotron frequency is larger than the wave frequency, 
\be
\om_ B \geq \om \rightarrow B \geq    350  \nu_9  \; {\rm G},
\label{Bfield}
\ee
then instead of large amplitude oscillations with $p_\perp$ given by Eq. (\ref {a0}) a particles experiences E-cross-B drift. 
Magnetic fields of this order do occur in early type stars \citep[\eg][]{2019MNRAS.489.5669P}.

\section{Predictions: mild DM variations and frequency-dependent activity window}
\label{prediction}
There is a number of  predictions. As a simple educated guess, we predict variations of the dispersion measure  within the observed window, and an increase of the  activity window at higher frequencies. First we give simple order-of-magnitude estimates, and then demonstrate that the reality is likely  to be more complicated.   \citep[RM and DM variations in wind of the binary pulsar PSR B1259 have been discussed by][]{1993ApJ...406..638K,1994MNRAS.268..430J,1995MNRAS.275..381M}. 


In  a homogeneous wind  a DM from   a  given point located distance $a$ from the central star and propagating with angle $\phi$
 with respect to the radial direction ($\phi$ is assumed to be fixed - we neglect possible plasma lensing effects),  is 
\be
DM (\phi) =  \int_a^\infty \frac{n_0 a^2}{x^2 +(x-a)^2 \tan^2 \phi} \frac{dx}{\cos \phi}   = \frac{\phi}{\sin \phi}DM_0,
\ee
where  
\be
DM_0 \approx  0.8  \; \dot{M}_{-7.5} \eta_{a,1.5}^{-1}  m_{\rm tot,1}^{-1/3} v_{w,8.5}^{-1}
\label{DM0}
\ee
is the dispersion measure for a radial ray. For larger $\phi$ the DM and the optical depth, see below,  increase as the light ray passes longer distance at higher densities. 

The wind can also produce  Rotation Measure (RM)  for the FRB pulse. To estimate the RM contribution of the wind, we scale the \Alfven\ velocity in the wind with the wind velocity, $v_A =\eta_A v_w$, where $\eta_A \sim 10^{-1}$.
We find
\be 
{\rm RM} = 7 \times 10^3   \left( \phi  \csc ^2 \phi -\cot \phi \right)\, \eta_{A,-1}   \eta_{a,1.5}^{-2}  \; \dot{M}_{-7.5}^{3/2} v_{w,8.5}^{-1/2}
\label{RM}
\ee
Narrow transparency window, $\phi \ll 1$ will further reduce the RM as the radio signal propagates mostly orthogonally to the toroidal \Bf\ in the wind (angular function in parenthesis $\propto \phi/3$ for $\phi \rightarrow 0$).

Though RM estimate (\ref{RM})  is fairly high,  it is a sensitive function of the assumed parameters:
 the uncertainties in the \Bf\ of the Main Sequence star  \citep{2019MNRAS.489.5669P}, wind dynamics (\eg\ ratio of the wind's  velocity to the \Alfven velocity in the wind), and the geometry of the \Bf\ in the wind. Observationally,  large RM $\sim 10^5  \rm{rad\, m}^{-2}$  was indeed observed in the case of 
 FRB 121102 \citep{2018Natur.553..182M}, while  FRB 180916.J0158+65 had small RM $\sim 144  \rm{rad\, m}^{-2}$ that may be completely due to  the Galactic foreground ISM \citep{2019ApJ...885L..24C}. For the Sun the RM is $\sim 10$ \citep{1994ApJ...434..773S}; we are not aware of RM measurements in the winds of massive stars.

\begin{figure}
 \begin{center}
 \includegraphics[width=.9\linewidth]{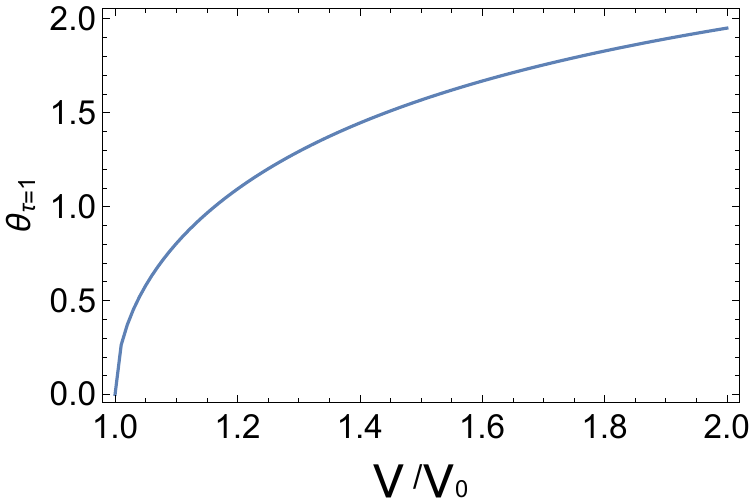}
\end{center}
\caption{Dependence of the  transparency angle $\phi$ on the observing frequency $\nu/\nu_0$ in idealized homogeneous wind. At the base frequency $\nu_0$ the radial (outward propagating) rays have $\tau=1$. At larger frequencies the rays at larger angles can escape, see also Fig. \protect \ref{fig:DM+FFa}}
\label{thetatau1}
\end{figure}

Similarly, in a homogeneous wind  an optical depth to a  given point (assuming isothermal wind)  is 
\ba &&
\tau (\phi) \propto \int_a^\infty  \left( \frac{n_0 a^2}{x^2 +(x-a)^2 \tan^2 \phi} \right)^2\frac{dx}{\cos \phi}
\nn &&
\tau (\phi) = \frac{3}{2} \left(\phi  \csc ^3(\phi )-\cot (\phi ) \csc (\phi )\right) \tau_0
\ea
where $\tau_0 = \tau(\phi=0)$ 
The optical depth for radial propagation depends on frequency, Eq. \ref{Mdot}
\be
 \tau_0 \propto \nu^{-2.1}
 \ee
Let $\nu_0$ be the frequency such that $\tau_0(\nu_0=1)$. At higher frequencies optical depth of 1 is reached at angles pictured in Fig. \ref{thetatau1}.

\begin{figure}
 \begin{center}
 \includegraphics[width=.97\linewidth]{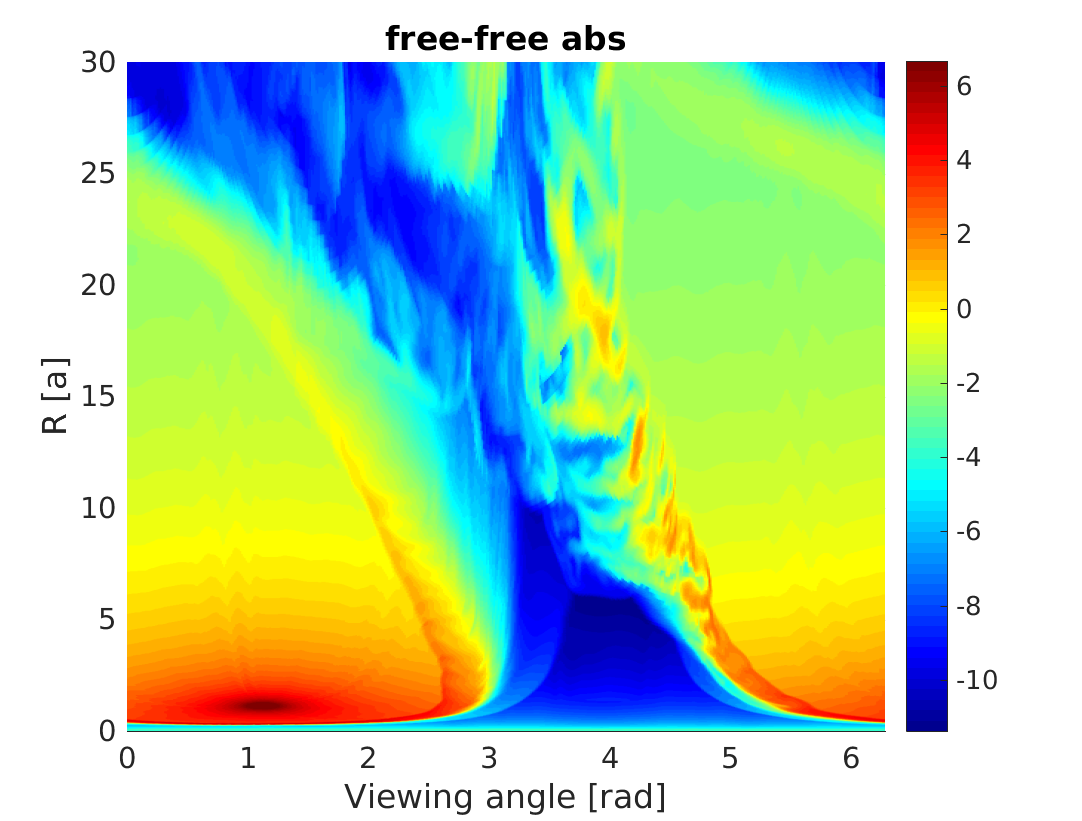}
\end{center}
\caption{Profiles of the differential absorption optical depth (color, depends on viewing angle and distance from compact object R) along different lines of sight  calculated near apastron orbital phase \citep[see details in][]{2015A&A...577A..89B}.  Low values (blue color) corresponds to transparency window. }
\label{fig:FFa_map}
\end{figure}

The above estimates assume idealized smooth wind. As numerical simulations demonstrate, Fig. \ref{WIndow}, the plasma around the bow shock is highly inhomogeneous.
The pulsar wind creates a tail cavity and an accumulation of dense material around the head.  These plasma ``wall'' will have an especially large effect on the free-free absorption (see Fig.\ref{fig:FFa_map}), 
since it depends on density squared. A plasma wall can be opaque to a broad range of frequencies, erasing a simple correlation between the active window and the observing frequency. 

\begin{figure}
 \begin{center}
 \includegraphics[width=.51\linewidth]{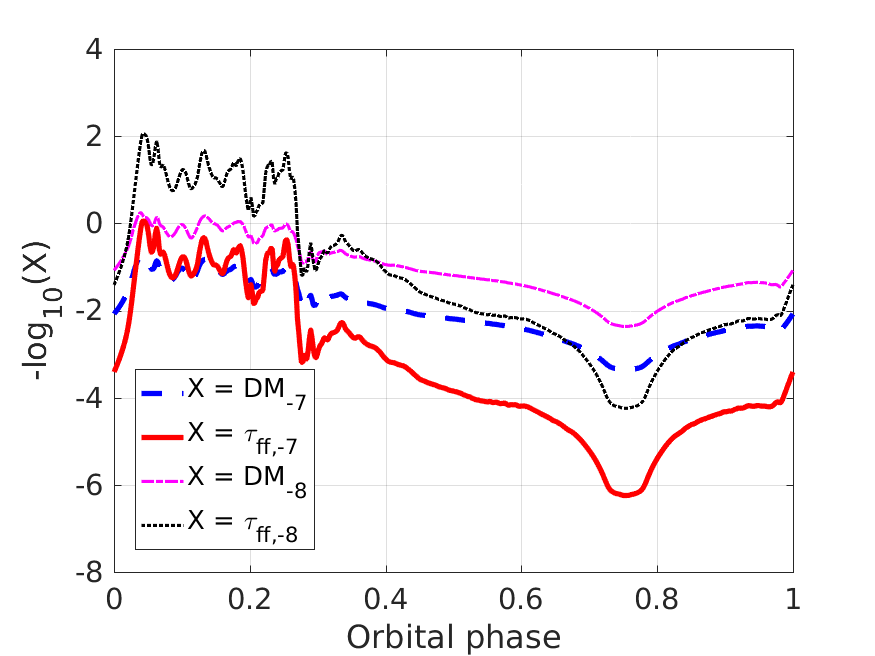}
 \includegraphics[width=.51\linewidth]{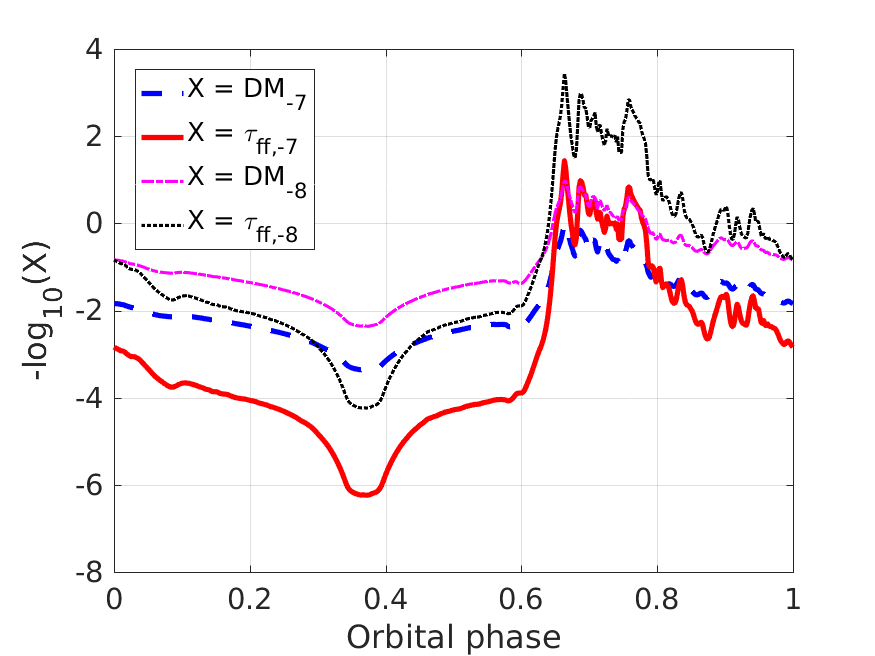}
 \includegraphics[width=.51\linewidth]{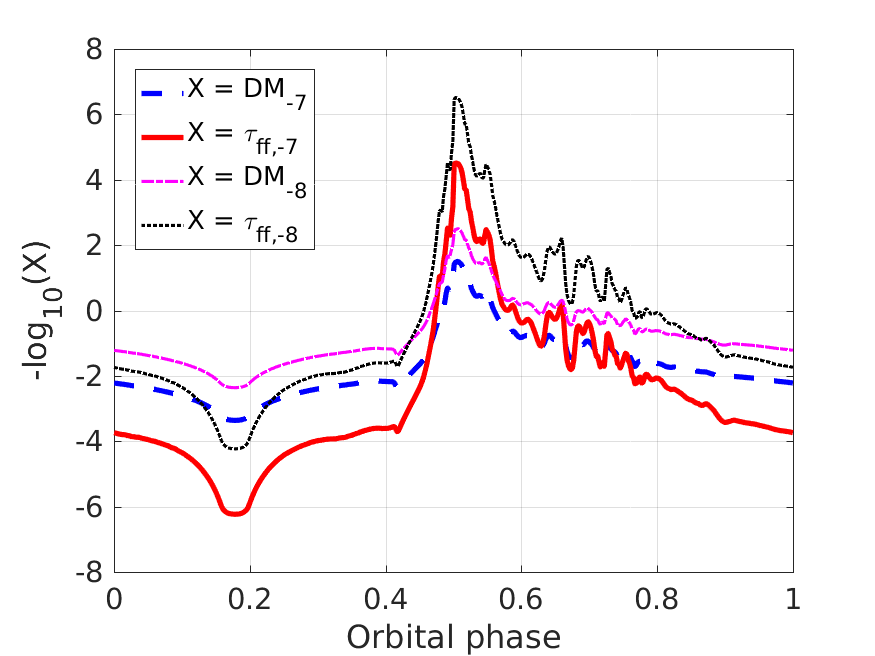}
\end{center}
\caption{Profiles of the DM (dashed blue $\dot{M}_{w} = 10^{-7} M_\odot$/yr and dot dashed magenta $\dot{M}_{w} = 10^{-8} M_\odot$/yr)  and absorption optical depth (solid red  
$\dot{M}_{w} = 10^{-7} M_\odot$/yr and doted black  $\dot{M}_{w} = 10^{-8} M_\odot$/yr) 
depending on orbital phases  calculated at   three different line of sight in equatorial plane (close to apastron - top, intermediate phase - center and close to periastron - bottom). Here we use the same hydro dynamical model as on Fig.~\ref{fig:FFa_map}, the orbital eccentricity   $e=0.24$. Large values of  $- \log \tau$
 (red curves) correspond to transparency window. }
\label{fig:DM+FFa}
\end{figure}

 We also calculated dispersion measure and free-free absorption using  the 3-dimensional relativistic hydrodynamical simulation by  \cite{2015A&A...577A..89B} of the 
  interaction of a massive stellar wind with the  pulsar wind. In their setup, the pulsar is moving on elliptic orbit with eccentricity $e=0.24$ and orbital period 3.4~days.  We  reprocessed the data for the orbital period of 16~days. \citep[Eccentricity for high mass binaries with compact object and orbital period about 15 days can vary from 0 till 0.85,][]{2011MNRAS.416.1556T,2017ApJ...846..170T}.
In  figure \ref{fig:DM+FFa}, we show the profiles of the integrated density, $\int n \propto  DM $ and  integrated density squared,
 $\int n^2 \propto  \tau $ 
  (the absorption optical depth)  along different lines of sight. 
  Fig. \ref{fig:DM+FFa} demonstrates that at each moment there is a narrow 
   transparency window with a duration from $\sim 0.1$ (close to apastron)  to $\sim 0.3$ (close to periastron) orbital period. Note that  a wide transparency window can be affected by the turbulent motion and can be chaotically transparent and opaque from orbit to orbit. 

Our analysis demonstrates that  the DM and free-free absorption show high correlation. The transparent window corresponds to a minimum of DM: this   can 
explain the small change of DM for repeating  FRBs. On the other hand, the variations of the DM within the transparency window are smaller by a factor of few than the overall variations, Fig. \ref{fig:FFa_map} bottom panel.

 In addition,  simulations show a sharp rise and decay in transparency near periastron phase, while    in direction of apastron such variations are more gradual, Fig. \ref{fig:DM+FFa} bottom panel. Thus we expect/predict orbital-dependent spectral evolution: burst near borders of transparency window could be bluer. Observational confirmation of such effect can be a smocking gun for our model.

\section{Discussion}

In this Letter we discuss a model for the periodicity observed in FRB 180916.J0158+65 by the CHIME telescope \citep{2020arXiv200110275T}. 
The working model is that the FRBs are produced by a {\NS} that orbits an OB primary. The periodicity 
 arises due to free-free absorption in the primary's (of the massive star's) wind. Thus, we argue that the  observed periodicity is a property of a particular system and is likely not essential 
for the FRB production. On the other hand, the association of an FRB with a compact stellar binary further strengthens the magnetospheric {\it loci} of FRBs, as argued by 
\citep{2013arXiv1307.4924P,2016MNRAS.462..941L,2019arXiv190103260L,2019arXiv190910409L} also \citep[see reviews by][]{2019ARA&A..57..417C,2019A&ARv..27....4P}. Note that  FRB 180916.J0158+65 also shows narrow emission bands drifting down in frequency - this is  naturally interpreted as plasma laser operating in \NS's \ms\ and showing radius-to-frequency mapping \citep{2019arXiv190910409L}.

{Observations of FRB~121102 by \citep{2020arXiv200303596R}   indicated periodicity of $P \sim 160$. We  Identify this periodicity with the orbital motion. The present model is applicable to the case of  FRB~121102 as well. 
Following equation~ (\ref{Mdot}) we can estimate the maximal orbital separation to absorb radio signal,  $2\times 10^{13}$~cm. This would   corresponds to the orbital period $\sim$  200 days.}

Most importantly, both  FRB 180916.J0158+65 \citep{2020arXiv200110275T}  and FRB~121102 \citep{2017Natur.541...58C,2017ApJ...834L...7T,2017ApJ...843L...8B}
are observed in  star-forming regions. This  is  consistent with the OB primary.

The model has three clear predictions: (i) mild variations of DM, Eq. (\ref{DM0}),  should be observed in the activity window; (ii) activity window should be broader at higher frequencies, Fig. \ref{thetatau1}; (iii) large RM should be measured in FRB 180916.J0158+65 \ref{prediction}. Importantly, in our model the binarity is not an intrinsic property of FRBs:  FRB~121102 and  FRB 180916.J0158+65 happen to be in binaries, other FRB sources can be isolated: in that case no large RMs are expected.

Interestingly, a new, unusual radiative mechanism operates when an FRB pulse enters the primary's wind: heating of  a narrow layer of plasma to relativistic energies by a radio beam. We are not aware of such processes in any other astrophysical system. Corresponding bursts of X-ray emission are likely to be too weak to be observable.

We constrain the primary to be  late O-type/early B-type star: earlier types have too powerful winds that remain heavily optically thick at the inferred orbital separation, while later types produce winds that are too weak - this runs contrary our interpretation of the small orbital activity window as a pointing to the momentum dominance of the primary's wind over the pulsar's.  
 Since the primary's momentum loss should over-power the pulsar's wind - the pulsar can be only mildly strong, with a spin-down luminosity $L_{sd} \sim 10^{37}$ erg s$^{-1}$; this values is somewhat larger than 
the spin-down power of  Galactic magnetars. 
 
 Periodic transparency can  also be achieved by having a  highly eccentric  orbit - then at an apastron the radio pulses would sample lower density plasma. We disfavor this scenario since it predicts a large  active window (since the binary would spend more time at large separations).

{Sharp  boundaries of the wind cavity can also lead to refractive effects. For isotropic FRBs, the reflection from the walls of the wind cavity will redistribute radio impulse in time so that its luminosity will by suppressed on factor $4\pi T_{FRB}c/\Omega_{cone} a \eta \sim 10^{-5} $. This will   make it difficult to detect the reflected signal.  On the other hand, if FRBs have narrow cone structure, then we could possibly  see several  echos formed by turbulence on cavity walls.}

 
 The present  model has  a number of similarities to binary  systems containing  pulsars. First,  binary pulsars PSR
1957 + 20 and  PSR 1744$ - $24A  shows periodic orbital-dependent  eclipses 
  \citep[limited to $\sim 10\%$  in phase in case of PSR
1957 + 20 and $\sim 50\%$ in case of PSR 1744$ - $24A][]{1990Natur.347..650L,1989ApJ...342..934R,1991A&A...241L..25R}. These are low mass binary, ablated by the NS's wind 
\citep{1988Natur.333..832P}, so the wind-wind interaction is dominated by the pulsar's wind, $\eta \geq 1$.  In that case a number of plasma effects, like variations of the 
DM and pulse delays, do show during the eclipse. 
Then, there is a number of $\gamma$-ray binaries that contain (or thought to contain)  {\NS}s 
\citep{2013A&ARv..21...64D,2016MNRAS.456L..64B,2017MNRAS.471L.150B,2018MNRAS.479.1320B,2018ApJ...867L..19A}

{In our view, the observed periodicity in FRB 180916.J0158+65 does fit with a general concept of {\NS}s' \mss\ being the {\it loci} of FRBs.  Identification of FBRs with \NSs\ leaves two possibilities for the energy source: rotational \citep[akin to Crab's giant pulses][]{popov06,2007MNRAS.381.1190L,2012ApJ...760...64M,2016MNRAS.462..941L},
 or magnetar-like magnetically powered emission \citep{2002ApJ...580L..65L,eichlerradiomagnetar}. 
  The present model   confirms that  FRBs are not rotationally powered \citep[since the present model requires that the pulsar wind should be  mild/weak][]{2017ApJ...838L..13L}. The magnetically powered model, radio emission generated in the magnetospheres of {\NS}s remains the most probable, in our view.}

{ Finally, can   the  FRBs'  radio emission be  induced   by  the interactions of the companions? Clearly not.   The orbital motionally-induced electric potential potential and luminosity  for   scaling (\ref{1})  estimate to 
\ba &&
\Phi_{orb} = e B a \beta_{orb} = 2 \times 10^{13}\, b_q\, \;{\rm  eV }
\nn &&
L_{orb} \sim ( \Phi_{orb}/e)^2 c = 2 \times 10^{31}  b_q ^2\,\; {\rm erg \; s}^{-1}, 
\label{Phi-E}
\ea
where we scaled the surface \Bf\ to quantum field, $b_q = B_{NS}/B_Q$, $B_Q = m_e^2 c^3/(e \hbar)$.
Estimates (\ref{Phi-E}) give  very weak power,  even for quantum-strong \Bfs, and small potential. We conclude that even in the tightest orbit  case of  geodetically induced precession, the powers expected due to the interaction of the binaries   are very  small. \citep[Note, that  binary pulsar PSR0737B that shows strong magnetospheric interaction is/was an  {\it exceptionally weak pulsar} - it cannot not be used as an model for FRBs,  the most powerful radio emitters.]{lyutikov04,lt05,2014MNRAS.441..690L}.  {\it Hence we conclude that 
binarity  is not a cause of FRB  emission. }}

\section*{Acknowledgments}

ML would like to acknowledge support by  NASA grant 80NSSC17K0757 and NSF grants  1903332 and 1908590.
We would like to thank Jason Hessels, Victoria Kaspi, Jonathan Katz, Wenbin Lu  and Elizaveta Ryspaeva for discussions and Yegor Lyutikov for help with the illustration.
\bibliographystyle{apj}
  \bibliography{../ApJ/BibTex,/Users/maxim/Home/Research/BibTex}

\appendix

\section{Unlikely alternative:  FRB variations due to geodetic precession}
\label{Unlikely}

Another possible source of variability is a the geodetic precession that makes the active region periodically aligned with the line of sight. For example, in the binary pulsar PSR 0737 the geodetic precession lead to the disappearance of the PSR 0737B \citep{Breton,2014MNRAS.441..690L}. To have a geodetic precession of only 16 days the required orbital size is 
\ba && 
\Omega_G = \left( \frac{2 \pi }{P} \right) ^{5/3}  \left( \frac{G M_\odot  }{c^3 } \right) ^{5/3} 
 \left( \frac{m_{MS} (4 m_{PSR}+3 m_{MS})}{2 (m_{PSR}+m_{MS})^{4/3}} \right)
\nn &&
a= 2 \times 10^9  \frac{(m_{MS} (4 m_{PSR}+3 m_{MS}))^{2/5} }{ (m_{PSR}+m_{MS})^{1/5}} \;  {\rm cm}
\label{1}
\ea
Gravitational decay time   is still  sufficiently long,
\be
t_{GW} =670  \frac{m_{MS}^{2/5}}{m_{PSR}}\; {\rm yr}
\ee
(where $m_{MS} \geq m_{PSR}$ was assumed). 

The system with separation of (\ref{1}) will be exceptional. Strong modifications of the magnetospheric properties may be expected in this case. For example,  setting separation equal to the light cylinder, the period would be 0.75 seconds.  Strong wind-\ms\ interactions are expected, similar to the case of the binary pulsar PSR0737A/B \cite{lyutikov04,lt05,2014MNRAS.441..690L}.

\end{document}